\documentclass{amsart}
\usepackage{graphicx}
\usepackage{subcaption}

\newtheorem{theorem}{\bf Theorem}[section]

\newtheorem{definition}{\bf Definition}[section]
\newtheorem{lemma}{\bf Lemma}[section]
\newtheorem{proposition}{\bf Proposition}[section]
\newtheorem{remark}{\bf Remark}[section]

\begin{document}

\title{On the Prandtl-Kolmogorov 1-equation model of turbulence}

\author{Kiera Kean$^{1}$, William Layton$^{2}$, and Michael Schneier$^{3}$}

\address{$^{1}$Department of Mathematics, University of Pittsburgh, Pittsburgh PA 15260, USA\\
$^{2}$Department of Mathematics, University of Pittsburgh, Pittsburgh PA 15260, USA \\
$^{3}$Department of Mathematics, University of Pittsburgh, Pittsburgh PA 15260, USA}



\email{kkh16@pitt.edu$^{1}$, wjl@pitt.edu$^{2}$, mhs64@pitt.edu$^{3}$}

\begin{abstract}
We prove an estimate of total (viscous plus modelled turbulent) energy
dissipation in general eddy viscosity models for shear flows. For general
eddy viscosity models, we show that the ratio of the near wall average
viscosity to the effective global viscosity is the key parameter. This
result is then applied to the 1-equation, URANS model of turbulence for
which this ratio depends on the specification of the turbulence length
scale. The model, which was derived by Prandtl in 1945, is a component of a
2-equation model derived by Kolmogorov in 1942 and is the core of many
unsteady, Reynolds averaged models for prediction of turbulent flows. Away
from walls, interpreting an early suggestion of Prandtl, we set%
\begin{equation*}
l=\sqrt{2}k^{+1/2}\tau, 
\end{equation*}%
where $\tau =$ selected time scale. In the near wall region analysis suggests replacing the traditional $l=0.41d$%
\ ($d=$ wall normal distance) with $l=0.41d\sqrt{d/L}$ giving, e.g.,%
\begin{equation*}
l=\min \left\{ \sqrt{2}k{}^{+1/2}\tau ,\text{ }0.41d\sqrt{\frac{d}{L}}%
\right\} . 
\end{equation*}%
This $l(\cdot )$ results in a simpler model with\ correct near wall
asymptotics. Its energy dissipation rate scales no larger than the
physically correct $O(U^{3}/L)$, balancing energy input with energy
dissipation.
\end{abstract}

\maketitle

\section{Introduction}

Predicting turbulent flows in practical settings means solving models
intended to predict averages of solutions of the Navier-Stokes (NS)
equations. Among a wide variety of approaches, summarized in Wilcox \cite%
{Wilcox}, eddy viscosity URANS (unsteady Reynolds Averaged NS) models are
used in many applications. Many are based on the 1-equation model of Prandtl 
\cite{P45} and Kolmogorov \cite{K42}, considered herein and given by%
\begin{gather}
v_{t}+v\cdot \nabla v-\nabla \cdot \left( \left[ 2\nu +\nu _{T}(\cdot )%
\right] \nabla ^{s}v\right) +\nabla p=f(x,y,z)\text{, }  \notag \\
\nabla \cdot v=0\text{, and }\nu _{T}=\mu l\sqrt{k},  \label{eq:1EqnModel} \\
k_{t}+v\cdot \nabla k-\nabla \cdot \left( \left[ \nu +\nu _{T}(\cdot )\right]
\nabla k\right) +\frac{1}{l}k\sqrt{k}=\nu _{T}(\cdot )|\nabla ^{s}v|^{2}%
\text{.}  \notag
\end{gather}%
Following for example \cite{MP} and \cite{Wilcox} p.37 eq. (3.9), $v$
approximates a finite time window average of the Navier-Stokes velocity $u$%
\begin{equation}
v(x,y,z,t)\simeq \overline{u}(x,y,z,t)=\frac{1}{\tau }\int_{t-\tau
}^{t}u(x,y,z,t^{\prime })dt^{\prime }.  \label{eq:Averaging}
\end{equation}%
The fluctuation is $u^{\prime }=u-\overline{u}$. Its associated turbulent
kinetic energy, approximated by the $k-$equation solution, is $k_{true}$ $=$ 
$\frac{1}{2}\overline{|u-\overline{u}|^{2}}(x,y,z,t)$. In (\ref{eq:1EqnModel}%
) $\nu $ is the kinematic viscosity, $p$ is a pressure, initial and boundary
conditions for $v$ and $k$ will be specified, $f$ is the body force (here $%
f=0$), $\nabla ^{s}v$ is the symmetric part of $\nabla v$ and $\nu
_{T}(\cdot )$\ is the eddy viscosity. The $k-$equation is derived, for
example, in \cite{CL} p.99, Section 4.4, \cite{D15}, \cite{MP} p.60, Section
5.3 or \cite{Pope} p.369, Section 10.3. The term $\nabla \cdot \left( \nu
\nabla k\right) $ in the $k-$equation is included in the model by some and
considered negligible by others.

The Kolmogorov-Prandtl relation is $\nu _{T}=\mu l\sqrt{k}$ where $\mu $ is
a calibration constant, typically $0.2$ to $0.6$, and often $0.55$, \cite{D15}
p. 114,\cite{Pope}. The turbulence length-scale $l=l(x,y,z,t)$ is specified
to complete the model. In current practice, $l(\cdot )$ varies from model to
model, subregion to subregion (requiring their locations, \cite{S15}) and
must be specified by the user; see \cite{Wilcox}, \cite{KKG17} for many
examples.

This lack of a simple, effective, and universal specification of $l(\cdot )$
is one disadvantage of $1-$equation models like (\ref{eq:1EqnModel}).
Another disadvantage, shared by many eddy viscosity models, is that model
dissipation often exceeds energy input and leads to lower Reynolds number
solutions. Herein we analyze a specification of $l$ with greater
universality and improved model dissipation%
\begin{gather}
l=\min \left\{ \sqrt{2}k{}^{+1/2}\tau ,\text{ }0.41d\sqrt{\frac{d}{L}}%
\right\} \text{, where}  \label{eq:MixingLength} \\
d=\text{wall distance},\tau =\text{averaging window, }L=\text{ global length
	scale.}  \notag
\end{gather}%
The main result herein, Theorem 4.1 Section 4, is that with (\ref%
{eq:MixingLength}) for shear flows, this over dissipation does not happen:
the model's energy dissipation rate is consistent with its energy input
rate. The effect of the minimum in (\ref{eq:MixingLength}) is to select $l=%
\sqrt{2}k{}^{+1/2}\tau $ in the flow's interior and the new value $l=0.41d%
\sqrt{d/L}$\ near walls. (Other realizations of this intent are
possible, e.g., (\ref{eq:TClength}).) The traditional value of the Von Karman
constant, $0.41$, is retained in (\ref{eq:MixingLength}). Prandtl \cite{P26}
described $l(\cdot )$ as "\textit{... the diameter of the masses of fluid
	moving as a whole in each individual case".} This diameter is constrained by
nearby walls leading to the classical $l=0.41d$ and here $0.41d\sqrt{d/L}$.
Prandtl also mentioned a second possibility, "...\textit{or again, as the
	distance traversed by a mass of this type before it becomes blended in with
	neighboring masses..."} This remark can be interpreted as $l=|u^{\prime
}(x,t)|\tau $, i.e., the \textit{distance a fluctuating eddy travels in one
	time unit.} As $|u^{\prime }|\simeq \sqrt{2}k^{+1/2}$, away from walls we
specify the kinematic relation 
\begin{equation}
l(\cdot )=\sqrt{2}k(\cdot )^{+1/2}\tau .
\end{equation}

\subsection{Justification of\textbf{\ }$l=0.41d\protect\sqrt{d/L}$\textbf{\ }%
}

The (dimensionally consistent) near wall $l=0.41d\sqrt{d/L}$ is a deviation
from accepted practice, so justification is necessary. The true turbulent
kinetic energy $k_{true}=\frac{1}{2}\overline{|u-\overline{u}|^{2}}$\ $%
\rightarrow 0$\ like $O(d^{2})$\ at walls. This rate implies that $k_{true}$%
\ satisfies 
\begin{equation*}
k_{true}=0\text{ and }\nabla k_{true}\cdot n=0\text{ at the wall.}
\end{equation*}%
The eddy viscosity should have a similar near wall behavior since, modulo
pressure terms, $\mu l\sqrt{k}\nabla ^{s}v\simeq u^{\prime }u^{\prime }$\ $%
\rightarrow 0$\ at walls like $O(d^{2}).$\ If $k_{true}$\ replaces $k$\ in $%
\nu _{T}(\cdot )$, then $\mu l\sqrt{k_{true}}\nabla ^{s}v=$\ $O(d^{2})$\ near
walls with $l=0.41d$. However, the solution to the $k-$equation satisfies
only one \ boundary condition, $k=0$\ at the wall. Thus, the solution to the $%
k$\ equation (intended to model $k_{true}$) has only 
\begin{equation*}
k=0\text{ at the wall, and }k(d)=\mathcal{O}(d)\text{ as the wall is
	approached.}
\end{equation*}%
This (incorrectly) implies $\mu l\sqrt{k}\nabla ^{s}v$\ $\rightarrow 0$\ at walls like $%
O(d^{+1.5})$\  when $l=$\ $0.41d$. This is one reason for
evaluations such as Pope \cite{Pope}\ p. 434 Section 11.7.2 that "\textit{\
	... the specification }$l=0.41y$\textit{\ is too large in the near wall
	region...}" as well as ad hoc addition of van Driest damping. The
modification $l=0.41d\sqrt{d/L}$\ in (\ref{eq:MixingLength}) ensures $\nu
_{T}(\cdot )=$\ $O(d^{2})$\ correctly in the model.

The question arises of why not simply specify $l(\cdot )=\sqrt{2}k(\cdot
)^{+1/2}\tau $\ as in \cite{LM18}. The positive results in \cite{LM18} were
for turbulence induced by a body force with $f(x)=0$\ on $\partial \Omega $
which excludes shear flows. The physical difference in the settings
(summarizing the introduction of Phillips \cite{P69}) is that in shear flows
the near wall region produces small scales which dominate $k_{true}$,
while when shear flows are excluded in \cite{LM18}, small scales are produced
only through the nonlinearity.

\subsection{Related work}

The energy dissipation rate is a fundamental statistic of turbulence, e.g., 
\cite{Pope}, \cite{V15}. Its balance with energy input rates, $\left\langle
\varepsilon \right\rangle =$ $\mathcal{O}(U^{3}/L)$, is observed in physical
experiments \cite{V15}. In 1992, Doering and Constantin \cite{DC92}
established a direct link between phenomenology and NSE predicted energy
dissipation through upper bounds consistent with the $\mathcal{O}(U^{3}/L)$\
rate. This work builds on \cite{B78}, \cite{H72} and has
developed in many important directions, e.g., \cite{DF02}, \cite{H72}, \cite%
{V15}, \cite{Wang97}, \cite{K97}, \cite{W00}.\ Remarkably, an $\mathcal{O}%
(U^{3}/L)$ lower bound has recently been proven in \cite{CP20} for
stochastically forced shear flow.

Model over-dissipation, producing a lower $\mathcal{R}e$\ flow, is due to
the action of turbulent viscosity terms on small scales generated by
breakdown of large scales through the nonlinearity or in the boundary layer. 
$\left\langle \varepsilon \right\rangle $ has been analyzed for some simpler
models, e.g., \cite{L02}, \cite{L16} (showing a dramatic difference between
shear and no shear cases), and \cite{P17}. The kinematic length scale $l=%
\sqrt{2}k^{+1/2}\tau $ occurred naturally in an ensemble algorithm in \cite%
{JL14b} and was highly developed by Teixeira and Cheinet \cite{TC01} and 
\cite{TC04} (see equation (7) on p. 2699), with near-wall transition to $l=0.41d$ by%
\begin{equation}
l=\theta (0.41d)+(1-\theta )\left( \sqrt{2}k^{+1/2}\tau \right) ,\text{ with 
}\theta =e^{-d/100}.  \label{eq:TClength}
\end{equation}%
The global specification $l=\sqrt{2}k^{+1/2}\tau $\ was proven in \cite{LM18}
not to over-dissipate with shear excluded (and boundary layers negligible).
This work leads to the problem considered herein to analyze shear/boundary
layer induced model dissipation.

Since $\tau $\ in (\ref{eq:Averaging}), (\ref{eq:MixingLength}) is user
supplied, it can be determined by the time scales required in an application
or related to a time step. The latter blurs the line between URANS and time
filtered large eddy simulation, \cite{P08}, as noted in the abstract of \cite%
{F10} "...\textit{most of the unsteady approaches ... can be regarded as a
	temporally filtered approach}." The time scale $\tau $\ can also be regarded
as a fundamental time scale of turbulence such as $\tau =k/\varepsilon ,$
e.g., \cite{SAA92}. Other natural choices of $\tau $ include $\tau \simeq
\delta /U,\delta =$ an estimate of layer-width \cite{TC01} and $\tau
=0.76/N,N=$ a selected-frequency, \cite{D76}.

\section{Shear Flow}

We analyze energy dissipation caused by the boundary layer  for shear
flow with zero body force, building on analysis in the pioneering paper \cite%
{DC92} and early work of Hopf \cite{H57}. Let the flow domain $\Omega
=(0,L)^{3}$and select $L-$periodic boundary conditions in $x,y$ and no-slip
at $z=0,z=L$. The wall is fixed at $z=0$ and the wall at $z=L$ slides with
velocity $(U,0,0)$:%
\begin{equation}
\begin{array}{cc}
Boundary & Conditions: \\ 
\text{moving top lid:} & v(x,y,L,t)=(U,0,0) \\ 
\text{fixed bottom wall:} & v(x,y,0,t)=0 \\ 
\text{periodic side walls:} & 
\begin{array}{c}
v(x+L,y,z,t)=v(x,y,z,t), \\ 
v(x,y+L,z,t)=v(x,y,z,t).%
\end{array}%
\end{array}
\label{eq:ShearBCs}
\end{equation}%
On this domain the wall normal distance is $d=\min \{z,L-z\}$. Since time
averages of the velocity satisfy the same shear boundary conditions as the
NSE solution, the correct boundary condition for $k(x,y,z,t)$ is%
\begin{equation*}
k(x,y,0,t)=k(x,y,L,t)=0\text{ and }L-\text{periodicity in }x,y\text{.}
\end{equation*}%
Since $k$ has homogeneous boundary conditions, non-zero initial conditions
must be specified; otherwise, if $k(x,y,z,0)=0$, then $k(x,y,z,t)\equiv 0$
thereafter.

\subsection{Notation and preliminaries}

The $L^{2}(\Omega )$ norm and the inner product are $\Vert \cdot \Vert $ and 
$(\cdot ,\cdot )$. The $L^{p}(\Omega )$ norms are $\Vert \cdot \Vert _{L^{p}}$%
. $C$ represents a generic positive constant independent of $\nu ,\mathcal{R}%
e$, and other model parameters.

\begin{definition}
	The finite and long time averages of a function $\phi (t)$\ are%
	\begin{equation*}
	\left\langle \phi \right\rangle _{T}=\frac{1}{T}\int_{0}^{T}\phi (t)dt\text{
		and }\left\langle \phi \right\rangle _{\infty }=\lim \sup_{T\rightarrow
		\infty }\left\langle \phi \right\rangle _{T}.
	\end{equation*}
\end{definition}

These satisfy $\left\langle \left\langle \phi \right\rangle _{\infty
}\right\rangle _{\infty }=\left\langle \phi \right\rangle _{\infty }$\ and 
\begin{equation}
\left\langle \phi \psi \right\rangle _{T}\leq \left\langle |\phi
|^{2}\right\rangle _{T}^{1/2}\left\langle |\psi |^{2}\right\rangle
_{T}^{1/2},\text{ }\left\langle \phi \psi \right\rangle _{\infty }\leq
\left\langle |\phi |^{2}\right\rangle _{\infty }^{1/2}\left\langle |\psi
|^{2}\right\rangle _{\infty }^{1/2}\text{\ }.  \label{eq:CSinTime}
\end{equation}%
A weak solution of the model momentum equation for shear flow problem
satisfies the initial condition and 
\begin{equation}
(v_{t},w)+([2\nu +\nu _{T}(\cdot )]\nabla ^{s}v,\nabla ^{s}w)+(v\cdot \nabla
v,w)=0  \label{varh:v}
\end{equation}%
for all test functions $w$, with $\nabla \cdot w=0$, $L-$periodic in $x$ and 
$y$ and $w(x,y,0,t)=0,w(x,y,L,t)=0$. If $\phi $ is a divergence free
function extending the shear boundary conditions (\ref{eq:ShearBCs}) into $%
\Omega $, formally taking the inner product with $w=v-\phi $ and expanding
gives%
\begin{gather*}
\frac{1}{2}\frac{d}{dt}||v||^{2}+\int_{\Omega }[2\nu +\,\nu _{T}(\cdot
)]|\,\nabla ^{s}{v}|^{2}dx= \\
=(v_{t},\phi )+\int_{\Omega }[2\nu +\,\nu _{T}(\cdot )]\nabla ^{s}{v}:\nabla
^{s}\phi dx+(v\cdot \nabla v,\phi ).
\end{gather*}

\begin{definition}
	The total energy dissipation rate (per unit volume) is%
	\begin{equation*}
	\varepsilon (v)=\frac{1}{|\Omega |}\int_{\Omega }[2\nu +\nu _{T}(\cdot
	)]|\nabla ^{s}v(x,t)|^{2}dx.
	\end{equation*}
\end{definition}

While a new $l(\cdot )$ gives a new model, existence of weak solutions to
models of this type is treated comprehensively in \cite{CL} and \cite{BM18}.
Herein, we assume that a weak solution of the model (\ref{eq:1EqnModel}), (%
\ref{eq:MixingLength}) with shear boundary conditions (\ref{eq:ShearBCs})
exists, $k\geq 0$ and solutions satisfy the energy inequality%
\begin{gather}
\frac{1}{2}\frac{d}{dt}||v||^{2}+\int_{\Omega }[2\nu +\,\nu _{T}(\cdot
)]|\,\nabla ^{s}{v}|^{2}dx\leq  \label{eq:EnergyINeq} \\
(v_{t},\phi )+\int_{\Omega }[2\nu +\,\nu _{T}(\cdot )]\nabla ^{s}{v}:\nabla
^{s}\phi dx+(v\cdot \nabla v,\phi ).  \notag
\end{gather}%
Using the energy inequality the appendix gives a proof of the following
bounds.

\begin{proposition}[Uniform Bounds]
	Consider the $1-$equation model (\ref{eq:1EqnModel}),\ (\ref{eq:MixingLength}%
	) with shear boundary conditions (\ref{eq:ShearBCs}). The following are
	uniformly bounded in $T$:%
	\begin{gather*}
	||v(T)||^{2},\int_{\Omega }k(T)dx,\int_{\Omega }\,\nu _{T}(\cdot ,T)dx, \\
	\left\langle \frac{1}{L^{3}}\int_{\Omega }|\,\nabla ^{s}{v}%
	|^{2}dx\right\rangle _{T},\left\langle \frac{1}{L^{3}}\int_{\Omega }\frac{1}{%
		l}k\sqrt{k}dx\right\rangle _{T}\text{ \ ,\ }\left\langle \frac{1}{L^{3}}%
	\int_{\Omega }[2\nu +\,\nu _{T}(\cdot )]|\,\nabla ^{s}{v}|^{2}dx\right%
	\rangle _{T}.
	\end{gather*}
\end{proposition}

\section{Energy dissipation in shear flows}

To formulate our first main result we first present a definition of the 
\textbf{effective viscosity} $\nu _{eff}$ ($\geq \nu $), the \textbf{average
	viscosity} in the boundary layer $\mathcal{S}_{\beta }$, and a few related
quantities. These are well defined due to the uniform bounds in Proposition
2.3.

\begin{definition}
	The \textit{effective viscosity} of solutions of (\ref{eq:1EqnModel}) under (%
	\ref{eq:ShearBCs}) is%
	\begin{equation*}
	\nu _{eff}:=\frac{\left\langle \frac{1}{|\Omega |}\int_{\Omega }[2\nu +\nu
		_{turb}(\cdot )]|\nabla ^{s}v|^{2}dx\right\rangle _{\infty }}{\left\langle 
		\frac{1}{|\Omega |}\int_{\Omega }|\nabla ^{s}v|^{2}dx\right\rangle _{\infty }%
	}.
	\end{equation*}%
	The large scale turnover time is $T^{\ast }=L/U$. The \textit{Reynolds number%
	} and \textit{effective Reynolds number} are%
	\begin{equation*}
	\mathcal{R}e=\frac{U\,L}{\nu }\text{ \ and\ }\mathcal{R}e_{eff}=\frac{U\,L}{%
		\nu _{eff}}.
	\end{equation*}%
	Let $\beta =\frac{1}{8}\mathcal{R}e_{eff}^{-1}$ and denote the region $%
	\mathcal{S}_{\beta }$ by%
	\begin{equation*}
	\mathcal{S}_{\beta }=\left\{ (x,y,z):0\leq x\leq L,0\leq y\leq L,(1-\beta
	)L<z<L\right\} .
	\end{equation*}%
	The \textit{average viscosity}, $\overline{\nu }$, in $\mathcal{S}_{\beta }$%
	\ is denoted%
	\begin{equation*}
	\overline{\nu }:=\left\langle \frac{1}{|\mathcal{S}_{\beta }|}\int_{\mathcal{%
			S}_{\beta }}[2\nu +\nu _{T}(\cdot )]dx\right\rangle _{\infty },\text{ where }%
	|\mathcal{S}_{\beta }|=\beta L^{3}\text{.}
	\end{equation*}
\end{definition}

Generally, the ratio of the effective and average viscosity is an
important statistic.

\begin{theorem}
	Suppose $\nu _{T}(\cdot )\geq 0$. Let $v$ be a weak solution of 
	\begin{equation}
	v_{t}+v\cdot \nabla v-\nabla \cdot \left( \left[ 2\nu +\nu _{T}(\cdot )%
	\right] \nabla ^{s}v\right) +\nabla p=0\text{, and }\nabla \cdot v=0  \notag
	\end{equation}%
	under (\ref{eq:ShearBCs}) satisfying the energy inequality (\ref%
	{eq:EnergyINeq}). Then, provided $\overline{\nu },\nu _{eff}$ are well
	defined,%
	\begin{equation*}
	\left\langle \varepsilon \right\rangle _{\infty }\leq \left\{ \frac{5}{2}+8%
	\frac{\overline{\nu }}{\nu _{eff}}\right\} \frac{U^{3}}{L}.
	\end{equation*}
\end{theorem}

\begin{remark}
	The multiplicative constants $5/2,8$ have not been optimized. Due to the
	problem symmetries and Galilean invariance, only the upper layer (near $z=L$%
	) needs to be monitored. For general shear flows, the average viscosity $%
	\overline{\nu }$ should be defined (and thus monitored) as the average over
	all (here upper and lower) boundary layers present.
\end{remark}

The proof begins with the background flow from Doering and Constantin \cite%
{DC92}, $\phi (z)=[\widetilde{\phi }(z),0,0]^{T}$\ where 
\begin{equation*}
\widetilde{\phi }(z)=\left\{ 
\begin{array}{cc}
0, & z\in \lbrack 0,L-\beta \,L] \\ 
\frac{U}{\beta \,L}(z-(L-\beta \,L)), & z\in \lbrack L-\beta \,L,L]%
\end{array}%
\right. \beta =\frac{1}{8}\mathcal{R}e_{eff}^{-1}.
\end{equation*}%
This function $\phi (z)$ is piecewise linear, continuous, divergence free
and satisfies the boundary conditions. We will need the following values of norms of $%
\phi $.

\begin{lemma}
	We have $\nabla \cdot \phi =0$ and%
	\begin{equation*}
	\begin{array}{cc}
	||\,\phi \,||_{L^{\infty }(\Omega )}=U, & ||\,\nabla \phi \,||_{L^{\infty
		}(\Omega )}=\frac{U}{\beta \,L},\text{ } \\ 
	||\,\phi \,||^{2}=\frac{1}{3}\,U^{2}\,\beta \,L^{3}, & \text{ }||\,\nabla
	\,\phi \,||^{2}=\frac{U^{2}\,L}{\beta }.%
	\end{array}%
	\end{equation*}
\end{lemma}

With this choice of $\phi ,$ time averaging the energy inequality (\ref%
{eq:EnergyINeq}) over $[0,T]$ and normalizing by $|\Omega |=L^{3}$ gives 
\begin{gather}
\frac{1}{2TL^{3}}||v(T)||^{2}+\left\langle \frac{1}{L^{3}}\int_{\Omega
}[2\nu +\,\nu _{T}(\cdot )]|\,\nabla ^{s}{v}|^{2}dx\right\rangle _{T}\leq
\label{eq:Step1} \\
\leq \frac{1}{2TL^{3}}||v(0)||^{2}+\frac{1}{TL^{3}}(v(T)-v(0),\phi
)+\left\langle \frac{1}{L^{3}}(v\cdot \nabla v,\phi )\right\rangle _{T}+ 
\notag \\
+\left\langle \frac{1}{L^{3}}\int_{\Omega }[2\nu +\,\nu _{T}(\cdot )]\nabla
^{s}{v}:\nabla ^{s}\phi dx\right\rangle _{T}  \notag.
\end{gather}%
Recall $\beta =\frac{1}{8}\mathcal{R}e_{eff}^{-1}.$ Due to Proposition 2.3, (%
\ref{eq:Step1}) can be written as%
\begin{equation}
\left\langle \varepsilon \right\rangle _{T}\leq \mathcal{O}(\frac{1}{T}%
)+\left\langle \frac{1}{L^{3}}(v\cdot \nabla v,\phi )\right\rangle
_{T}+\left\langle \frac{1}{L^{3}}\int_{\Omega }[2\nu +\,\nu _{T}(\cdot
)]\nabla ^{s}{v}:\nabla ^{s}\phi dx\right\rangle _{T}
\end{equation}%
The right-hand side (RHS) has two terms shared by the NSE, $(v\cdot \nabla v,\phi )$\ and $%
\int 2\nu \nabla ^{s}{v}:\nabla ^{s}\phi dx$. The main issue is thus the
third term, $\int \,\nu _{T}(\cdot )\nabla ^{s}{v}:\nabla ^{s}\phi dx$.
Before treating that we recall the analysis of Doering and Constantine \cite%
{DC92} and Wang \cite{Wang97} for the first two. For the nonlinear term $%
\left\langle \frac{1}{L^{3}}(v\cdot \nabla v,\phi )\right\rangle _{T}$,
denoted $NLT$, we have%
\begin{gather*}
NLT=\left\langle \frac{1}{L^{3}}(v\cdot \nabla v,\phi )\right\rangle
_{T}=\left\langle \frac{1}{L^{3}}([v-\phi ]\cdot \nabla v,\phi
)\right\rangle _{T}+\left\langle \frac{1}{L^{3}}(\phi \cdot \nabla v,\phi
)\right\rangle _{T} \\
\leq \left\langle \frac{1}{L^{3}}\int_{\mathcal{S}_{\beta }}|v-\phi ||\nabla
v||\phi |+|\phi |^{2}|\nabla v|dx\right\rangle _{T} \\
\leq \frac{1}{L^{3}}\left\langle \left\Vert \frac{v-\phi }{L-z}\right\Vert
_{L^{2}(\mathcal{S}_{\beta })}||\nabla v||_{L^{2}(\mathcal{S}_{\beta
	})}||(L-z)\phi ||_{L^{\infty }(\mathcal{S}_{\beta })}+||\phi ||_{L^{\infty }(%
	\mathcal{S}_{\beta })}^{2}||\nabla v||_{L^{1}(\mathcal{S}_{\beta
	})}\right\rangle _{T}.
\end{gather*}%
On the RHS, $||\phi ||_{L^{\infty }(\mathcal{S}_{\beta })}^{2}=U^{2}$. We
calculate $||(L-z)\phi ||_{L^{\infty }(\mathcal{S}_{\beta })}=\frac{1}{4}%
\beta LU.$ Since $v-\phi $\ vanishes on $\partial \mathcal{S}_{\beta }$,
Hardy's inequality, the triangle inequality and a calculation imply 
\begin{eqnarray*}
	\left\Vert \frac{v-\phi }{L-z}\right\Vert _{L^{2}(\mathcal{S}_{\beta })}
	&\leq &2\left\Vert \nabla (v-\phi )\right\Vert _{L^{2}(\mathcal{S}_{\beta
		})}\leq 2\left\Vert \nabla v\right\Vert _{L^{2}(\mathcal{S}_{\beta
		})}+2\left\Vert \nabla \phi \right\Vert _{L^{2}(\mathcal{S}_{\beta })} \\
	&\leq &2\left\Vert \nabla v\right\Vert _{L^{2}(\mathcal{S}_{\beta })}+2U%
	\sqrt{\frac{L}{\beta }}.
\end{eqnarray*}%
Thus we have the estimate%
\begin{equation}
NLT\leq \frac{\beta LU}{4}\frac{1}{L^{3}}\left\langle 2||\nabla v||_{L^{2}(%
	\mathcal{S}_{\beta })}^{2}+2U\sqrt{\frac{L}{\beta }}||v||_{L^{2}(\mathcal{S}%
	_{\beta })}\right\rangle _{T}+\frac{U^{2}}{L^{3}}\left\langle ||\nabla
v||_{L^{1}(\mathcal{S}_{\beta })}\right\rangle _{T}.  \label{eq:NLTest}
\end{equation}%
For the last term on the RHS, H\"{o}lders inequality in space then in time
implies%
\begin{eqnarray*}
	\frac{U^{2}}{L^{3}}\left\langle ||\nabla v||_{L^{1}(\mathcal{S}_{\beta
		})}\right\rangle _{T} &=&\frac{U^{2}}{L^{3}}\left\langle \int_{\mathcal{S}%
		_{\beta }}|\nabla v|\cdot 1dx\right\rangle _{T}\leq \frac{U^{2}}{L^{3}}%
	\left\langle \sqrt{\int_{\mathcal{S}_{\beta }}|\nabla v|^{2}dx}\sqrt{\beta
		L^{3}}\right\rangle _{T} \\
	&\leq &\frac{U^{2}\sqrt{\beta }}{L^{3/2}}\left\langle \sqrt{\int_{\mathcal{S}%
			_{\beta }}|\nabla v|^{2}dx}\right\rangle _{T}\leq \frac{U^{2}\sqrt{\beta }}{%
		L^{3/2}}\left\langle \int_{\mathcal{S}_{\beta }}|\nabla
	v|^{2}dx\right\rangle _{T}^{1/2}.
\end{eqnarray*}%
Increase the integral's domain from $\mathcal{S}_{\beta }$ to $\Omega $, use
(as $\nabla \cdot v=0$) $||\nabla v||^{2}=2||\nabla ^{s}v||^{2}$ and $\beta =%
\frac{1}{8}\mathcal{R}e_{eff}^{-1}.$ Rearranging and using the
arithmetic-geometric inequality gives%
\begin{gather*}
\frac{U^{2}}{L^{3}}\left\langle ||\nabla v||_{L^{1}(\mathcal{S}_{\beta
	})}\right\rangle _{T}\leq U^{2}\sqrt{\beta }\left\langle \frac{1}{L^{3}}%
\int_{\Omega }2|\nabla ^{s}v|^{2}dx\right\rangle _{T}^{1/2}\leq \\
\leq U^{2}\sqrt{\frac{2}{8}\frac{1}{LU}}\left\langle \frac{1}{L^{3}}%
\int_{\Omega }\nu _{eff}|\nabla ^{s}v|^{2}dx\right\rangle _{T}^{1/2}\leq
\left( \frac{U^{3}}{L}\right) ^{1/2}\frac{1}{2}\left\langle \frac{1}{L^{3}}%
\int_{\Omega }\nu _{eff}|\nabla ^{s}v|^{2}dx\right\rangle _{T}^{1/2} \\
\leq \frac{1}{2}\frac{U^{3}}{L}+\frac{1}{8}\left\langle \frac{1}{L^{3}}%
\int_{\Omega }\nu _{eff}|\nabla ^{s}v|^{2}dx\right\rangle _{T}.
\end{gather*}%
Similar manipulations yield%
\begin{gather*}
\frac{1}{4}\beta LU\frac{1}{L^{3}}\left\langle 2U\sqrt{\frac{L}{\beta }}%
||v||_{L^{2}(\mathcal{S}_{\beta })}\right\rangle _{T}\leq \frac{1}{2}\beta
LU\left\langle \frac{1}{L^{3}}||\nabla v||_{L^{2}(\mathcal{S}_{\beta
	})}^{2}\right\rangle _{T}+\frac{1}{8}\frac{U^{3}}{L} \\
\leq \frac{1}{8}\left\langle \frac{1}{L^{3}}\nu _{eff}||\nabla
^{s}v||_{L^{2}(\mathcal{S}_{\beta })}^{2}\right\rangle _{T}+\frac{1}{8}\frac{%
	U^{3}}{L}.
\end{gather*}%
Using the last two estimates in the $NLT$ upper bound (\ref{eq:NLTest}), we
obtain%
\begin{equation*}
NLT\leq 2\beta \frac{LU}{\nu _{eff}}\left\langle \frac{1}{L^{3}}\nu
_{eff}||\nabla ^{s}v||_{L^{2}(\mathcal{S}_{\beta })}^{2}\right\rangle _{T}+%
\frac{5}{8}\frac{U^{3}}{L}.
\end{equation*}%
Thus,%
\begin{eqnarray*}
	\left\langle \varepsilon \right\rangle _{T} &\leq &\mathcal{O}(\frac{1}{T})+%
	\frac{1}{4}\left\langle \frac{1}{L^{3}}\nu _{eff}||\nabla
	^{s}v||_{L^{2}(\Omega )}^{2}\right\rangle _{T}+\frac{5}{8}\frac{U^{3}}{L}+ \\
	&&+\left\langle \frac{1}{L^{3}}\int_{\Omega }[2\nu +\,\nu _{T}(\cdot
	)]\nabla ^{s}{v}:\nabla ^{s}\phi dx\right\rangle _{T}.
\end{eqnarray*}%
Consider now the last term on the RHS. Since $\phi $ is zero off $\mathcal{S}%
_{\beta }$,
\begin{gather*}
\left\langle \frac{1}{L^{3}}\int_{\Omega }[2\nu +\,\nu _{T}(\cdot )]\nabla
^{s}{v}:\nabla ^{s}\phi dx\right\rangle _{T}=\left\langle \frac{1}{L^{3}}%
\int_{\mathcal{S}_{\beta }}[2\nu +\,\nu _{T}(\cdot )]\nabla ^{s}{v}:\nabla
^{s}\phi dx\right\rangle _{T} \\
\leq \frac{1}{2}\left\langle \varepsilon \right\rangle _{T}+\frac{1}{2}%
\left\langle \frac{1}{L^{3}}\int_{\mathcal{S}_{\beta }}[2\nu +\,\nu
_{T}(\cdot )]\left( \frac{U}{\beta L}\right) ^{2}dx\right\rangle _{T} \\
\leq \frac{1}{2}\left\langle \varepsilon \right\rangle _{T}+\frac{1}{2}%
\left( \frac{U}{\beta L}\right) ^{2}\beta \left\langle \frac{1}{\beta L^{3}}%
\int_{\mathcal{S}_{\beta }}[2\nu +\,\nu _{T}(\cdot )]dx\right\rangle _{T}.
\end{gather*}%
Thus%
\begin{eqnarray*}
	\frac{1}{2}\left\langle \varepsilon \right\rangle _{T} &\leq &\mathcal{O}(%
	\frac{1}{T})+\frac{1}{4}\left\langle \frac{1}{L^{3}}\nu _{eff}||\nabla
	^{s}v||_{L^{2}(\Omega )}^{2}\right\rangle _{T}+ \\
	&&+\frac{5}{8}\frac{U^{3}}{L}+\frac{\beta }{2}\left( \frac{U}{\beta L}%
	\right) ^{2}\left\langle \frac{1}{\beta L^{3}}\int_{\mathcal{S}_{\beta
	}}2\nu +\,\nu _{T}(\cdot )dx\right\rangle _{T}.
\end{eqnarray*}%
As $T\rightarrow \infty $%
\begin{equation*}
\left\langle \frac{1}{\beta L^{3}}\int_{\mathcal{S}_{\beta }}2\nu +\,\nu
_{T}(\cdot )dx\right\rangle _{T}\rightarrow \overline{\nu }\text{ and }%
\left\langle \frac{1}{L^{3}}\nu _{eff}||\nabla ^{s}v||_{L^{2}(\Omega
	)}^{2}\right\rangle _{T}\rightarrow \left\langle \varepsilon \right\rangle
_{\infty }.
\end{equation*}%
Thus,%
\begin{equation*}
\left( \frac{1}{2}-2\beta \mathcal{R}e_{eff}\right) \left\langle \varepsilon
\right\rangle _{\infty }\leq \frac{5}{8}\frac{U^{3}}{L}+\frac{1}{2}\left( 
\frac{U}{\beta L}\right) ^{2}\beta \overline{\nu }\leq \left[ \frac{5}{8}+%
\frac{1}{2\beta }\mathcal{R}e_{eff}^{-1}\frac{\overline{\nu }}{\nu _{eff}}%
\right] \frac{U^{3}}{L}.
\end{equation*}%
The choice $\beta =\frac{1}{8}\mathcal{R}e_{eff}^{-1}$\ implies $2\beta 
\mathcal{R}e_{eff}=1/4$, completing the proof since%
\begin{equation*}
\left\langle \varepsilon \right\rangle _{\infty }\leq \frac{5}{2}\frac{U^{3}%
}{L}+\frac{1}{2}\left( \frac{U}{\beta L}\right) ^{2}\beta \overline{\nu }=%
\left[ \frac{5}{2}+8\frac{\overline{\nu }}{\nu _{eff}}\right] \frac{U^{3}}{L}%
.
\end{equation*}

\section{Application to a 1-equation URANS model}

We now apply Theorem 3.2 to (\ref{eq:1EqnModel}), (\ref%
{eq:MixingLength}). The main work will be in estimating $\frac{\overline{\nu 
}}{\nu _{eff}}$\ .

\begin{theorem}
	Let $v$ be a weak solution of (\ref{eq:1EqnModel}), (\ref{eq:MixingLength})
	under (\ref{eq:ShearBCs}) satisfying the energy inequality (\ref%
	{eq:EnergyINeq}). We have%
	\begin{equation*}
	\left\langle \varepsilon \right\rangle _{\infty }\leq \left[ 5+32\frac{\nu }{%
		\nu _{eff}}+\left( \frac{0.41^{2}\sqrt[2]{2}\mu ^{2}}{4}\right) \frac{\tau }{%
		T^{\ast }}\right] \frac{U^{3}}{L}.
	\end{equation*}
\end{theorem}

\textbf{Remark}. We note that $\frac{\nu }{\nu _{eff}}\leq 1$ (and possibly $%
<<1$) and for $\mu =0.55$, $0.41^{2}\sqrt[2]{2}\mu ^{2}/4\simeq 0.017978.$

\textbf{proof}. The upper bound $l\leq 0.41d\sqrt{d/L}$ is used in the
boundary layer region to estimate $\overline{\nu }$ as follows 
\begin{gather}
\overline{\nu }=\left\langle \frac{1}{\beta L^{3}}\int_{\mathcal{S}_{\beta
}}2\nu +\nu _{T}(\cdot )dx\right\rangle _{\infty }\leq 2\nu +\left\langle 
\frac{1}{\beta L^{3}}\int_{\mathcal{S}_{\beta }}\mu \left( 0.41d\sqrt{\frac{d%
	}{L}}\right) k^{\frac{1}{2}}dx\right\rangle _{\infty }  \notag \\
\leq 2\nu +0.41\mu \frac{1}{L^{+1/2}}\frac{1}{\beta L^{3}}\left\langle \int_{%
	\mathcal{S}_{\beta }}\left( L-z\right) ^{+3/2}k^{+1/2}dx\right\rangle
_{\infty }  \notag \\
\leq 2\nu +0.41\mu \frac{1}{L^{+1/2}}\frac{1}{\beta L^{3}}\left\langle \sqrt{%
	\int_{\mathcal{S}_{\beta }}\left( L-z\right) ^{3}dx}\sqrt{\int_{\mathcal{S}%
		_{\beta }}kdx}\right\rangle _{\infty }  \notag \\
\leq 2\nu +\frac{0.41\mu }{2}\frac{1}{L^{+1/2}}\beta \left\langle \sqrt{%
	\int_{\mathcal{S}_{\beta }}kdx}\right\rangle _{\infty },\text{ hence}  \notag
\\
\text{ }\overline{\nu }\leq 2\nu +\frac{0.41\mu }{2}L\beta \sqrt{%
	\left\langle \frac{1}{L^{3}}\int_{\Omega }kdx\right\rangle _{\infty }}.
\label{eq:nuBARestimate}
\end{gather}%
Next use the $k-$equation to estimate $\int kdx$. We have%
\begin{equation}
\int_{\Omega }k_{t}dx+\int_{\Omega }\frac{1}{l}k\sqrt{k}dx=\int_{\Omega }\nu
_{T}(\cdot )|\nabla ^{s}v|dx.  \label{eq:IntegratedKequation}
\end{equation}%
By the choice of $l$, $\frac{1}{l}k\sqrt{k}$ is bounded below by $\frac{1}{%
	\sqrt{2}\tau }k$\ because%
\begin{equation*}
\frac{1}{l}k\sqrt{k}=\max \left\{ \frac{1}{\sqrt{2}\tau },\text{ }\frac{%
	\sqrt{k}}{0.41d\sqrt{\frac{d}{L}}}\right\} k\geq \frac{1}{\sqrt{2}\tau }k.
\end{equation*}%
The long time averaging of $\int k_{t}dx$ is zero. Since $\frac{1}{\sqrt{2}%
	\tau }k\leq \frac{1}{l}k\sqrt{k},$ we have%
\begin{equation*}
\frac{1}{\sqrt{2}\tau }\left\langle \frac{1}{|\Omega |}\int_{\Omega
}kdx\right\rangle _{\infty }\leq \left\langle \frac{1}{|\Omega |}%
\int_{\Omega }\frac{1}{l}k\sqrt{k}dx\right\rangle _{\infty }=\left\langle 
\frac{1}{|\Omega |}\int_{\Omega }\nu _{T}(\cdot )|\nabla
^{s}v|^{2}dx\right\rangle _{\infty }=\left\langle \varepsilon \right\rangle
_{\infty }.
\end{equation*}%
Thus, $\left\langle \frac{1}{|\Omega |}\int_{\Omega }kdx\right\rangle
_{\infty }\leq \sqrt{2}\tau \left\langle \varepsilon \right\rangle _{\infty
} $. Using this upper estimate in (\ref{eq:nuBARestimate}) we obtain%
\begin{equation*}
\overline{\nu }\leq 2\nu +\frac{0.41\mu }{2}L\beta \sqrt{\left\langle \frac{1%
	}{L^{3}}\int_{\Omega }kdx\right\rangle _{\infty }}\leq 2\nu +\frac{0.41\sqrt[%
	4]{2}\mu }{2}L\beta \tau ^{1/2}\sqrt{\left\langle \varepsilon \right\rangle
	_{\infty }}.
\end{equation*}%
Divide by $\mathcal{\nu }_{eff}$, use $T^{\ast }=L/U,\beta =\frac{1}{8}%
\mathcal{R}e_{eff}^{-1}$ and rearrange. This gives%
\begin{equation*}
\frac{\overline{\nu }}{\mathcal{\nu }_{eff}}\leq 2\frac{\nu }{\mathcal{\nu }%
	_{eff}}+\frac{0.41\sqrt[4]{2}\mu }{2}\frac{1}{8}\frac{L^{1/2}}{U^{3/2}}\sqrt{%
	\frac{\tau }{T^{\ast }}}\sqrt{\left\langle \varepsilon \right\rangle
	_{\infty }}.
\end{equation*}%
Using this estimate in Theorem 3.2 gives%
\begin{equation*}
\left\langle \varepsilon \right\rangle _{\infty }\leq \left[ \frac{5}{2}+16%
\frac{\nu }{\nu _{eff}}\right] \frac{U^{3}}{L}+\left[ \frac{0.41\sqrt[4]{2}%
	\mu }{2}\sqrt{\frac{\tau }{T^{\ast }}}\sqrt{\frac{U^{3}}{L}}\right] \sqrt{%
	\left\langle \varepsilon \right\rangle _{\infty }}.
\end{equation*}%
The arithmetic-geometric mean inequality then completes the proof:%
\begin{equation*}
\left\langle \varepsilon \right\rangle _{\infty }\leq \left[ 5+32\frac{\nu }{%
	\nu _{eff}}+\frac{0.41^{2}\sqrt[2]{2}\mu ^{2}}{4}\frac{\tau }{T^{\ast }}%
\right] \frac{U^{3}}{L}.
\end{equation*}

\section{A Numerical Illustration}

This section provides a computational illustration of the theoretical
results for the model. The results of the computations are consistent with
the theoretical predictions. The results were obtained on a workstation with
a program developed with the FEniCS software suite \cite{FENICS}. The code can be found on GitHub at https://github.com/kierakean/1eqnRANS-FEM.

\subsection{Problem Setting}

\begin{figure}[h]
	\centering
	\begin{subfigure}[b]{0.33\linewidth}
		\includegraphics[width=\linewidth]{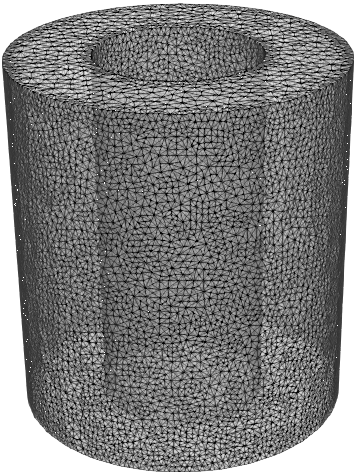}
		\caption{The Domain $\Omega$.}
	\end{subfigure}
	\begin{subfigure}[b]{0.33\linewidth}\label{meshcross}
		\includegraphics[width=\linewidth]{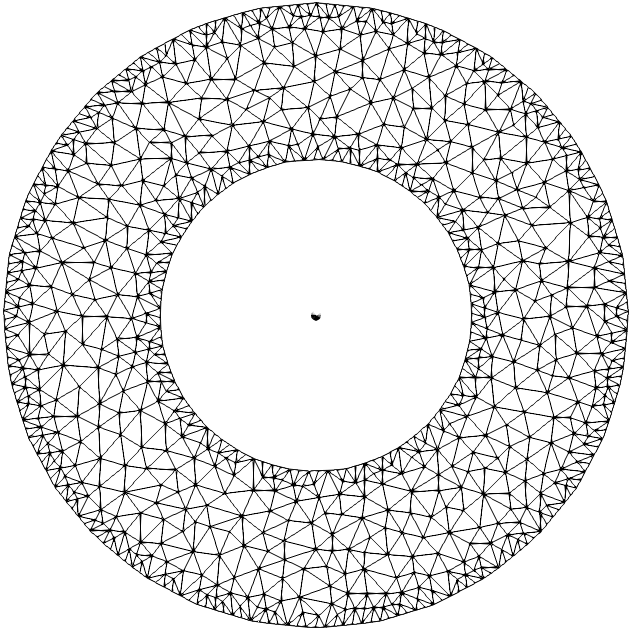}
		\caption{The mesh viewed from the top.}
	\end{subfigure}
	\caption{{The unstructured mesh used in the numerical experiments.}} \label{domainmesh}
\end{figure}

We examined the classical Taylor-Couette flow between
counter-rotating cylinders for rotations well above, e.g. \cite{GLS16},
those yielding stable patterns, \cite{T23}. The domain is given by 
\begin{equation*}
\Omega =\{(x,y,z):r_{inner}^{2}\leq x^{2}+y^{2}\leq r_{outer}^{2},0\leq
z\leq z_{max}\},
\end{equation*}%
with $r_{inner}=.5,r_{outer}=1,z_{max}=2.2$. Figure \ref{domainmesh}. (a)
depicts the domain $\Omega$.


 We imposed periodic boundary conditions in the $z$
direction. The outer cylinder was held fixed and the flow was driven by the
rotation of the inner cylinder. The angular velocity of the inner cylinder, $%
\omega _{inner}$ was smoothly increased from zero at $T=0$ to $\omega
_{inner}=4$ at $T=5$. Plots of flow statistics indicated that statistical
equilibrium was reached around $T=20$ so we give snapshots below at $T=30$. We
chose final time $T=40$ and time averaged over $20\leq t\leq 40$. The time
scale was chosen to be $\tau =0.1$.

\textbf{Initialization.} The model is turned on with a non-zero $k(x$,$5)$
at $T=5$ when the inner cylinder has been spun up to its full angular
velocity. We use a $k$ initialization standard for turbulent flow in a
square duct, Wilcox \cite{Wilcox}, given by%
\begin{equation*}
k(x,5)=1.5|v(x,5)|^{2}I^{2},\text{ where }I=\text{ turbulence intensity }%
\simeq 0.16\mathcal{R}e^{-1/8}.
\end{equation*}

\textbf{The mesh}. We used an unstructured mesh that was refined around the inner and outer boundaries,
as can be seen from the top of the mesh in Figure \ref{domainmesh} (b). We did preliminary tests at
Reynolds number $\mathcal{R}e=1000$ by refining the mesh until $\left\langle \varepsilon \right\rangle $\
was unchanged on three successive refinements. These parameters yielded a Taylor
number of%
\begin{equation*}
Ta:=\frac{\omega ^{2}r_{inner}(r_{outer}-r_{inner})^{3}}{\nu ^{2}}=10^{6}.
\end{equation*}
We then did all reported tests on the coarsest mesh that produced the same
value of $\left\langle \varepsilon \right\rangle $. 

\begin{figure}[h]
	\centering
	\begin{subfigure}[b]{0.33\linewidth}
		\includegraphics[width=\linewidth]{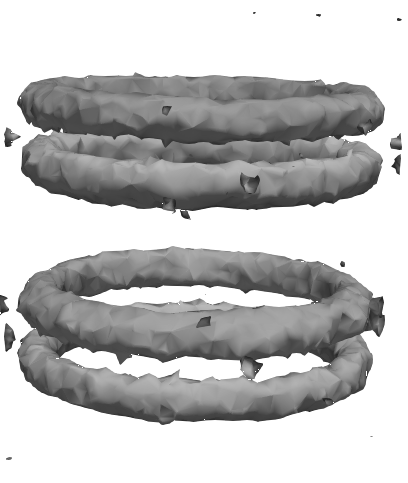}
		\caption{$\nu = .003:$ Clear coherent vortices.}
	\end{subfigure}
	\begin{subfigure}[b]{0.33\linewidth}
		\includegraphics[width=\linewidth]{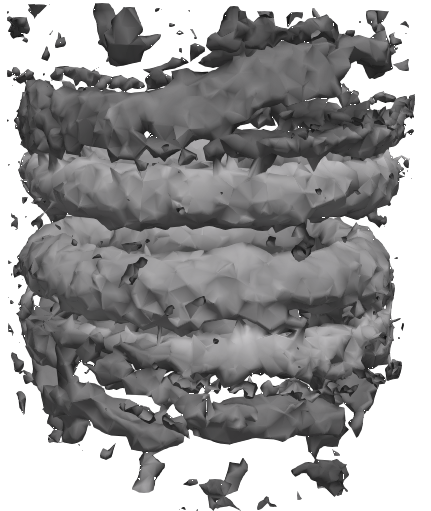}
		\caption{$\nu = .0005$: Vortices not axisymmetric.}
	\end{subfigure}
	\caption{Q-criterion at T = 30.}\label{qcrit}
\end{figure}

Tests were run with varying Reynolds numbers by varying the viscosity $\nu $ from $3 \times 10^{-3}$
to $5 \times 10^{-4}$ ($Ta\simeq $ $10^{10}$ to $2.5\times 10^{17}$). Persistent
vortices, marked by the Q-criterion, are plotted for two Reynolds numbers in
Figure \ref{qcrit}.

We used the $P^{2}-P^{1}$ Taylor-Hood element pair. The velocity space, $%
X_{h}$ and pressure space, $Q_{h}$ had $947,802$ and $44,585$ degrees of
freedom, respectively. We used The timestepping scheme backward Euler plus time
filter from \cite{GL18} for the momentum and continuity equation.
The added time filter increased accuracy and reduced numerical dissipation
making the calculated $\left\langle \varepsilon \right\rangle $\ more
accurate. We used Backward Euler for the $k$ equation. This choice smoothed
the $k(x,t)$ evolution and reduced solver issues. We took $%
\Delta t=1e-2$ and ran the simulation from $T=0$ to $T=40$.


\subsection{Energy Dissipation Rate}

\begin{figure}[h]
	\centering
	\includegraphics[width=0.5\linewidth]{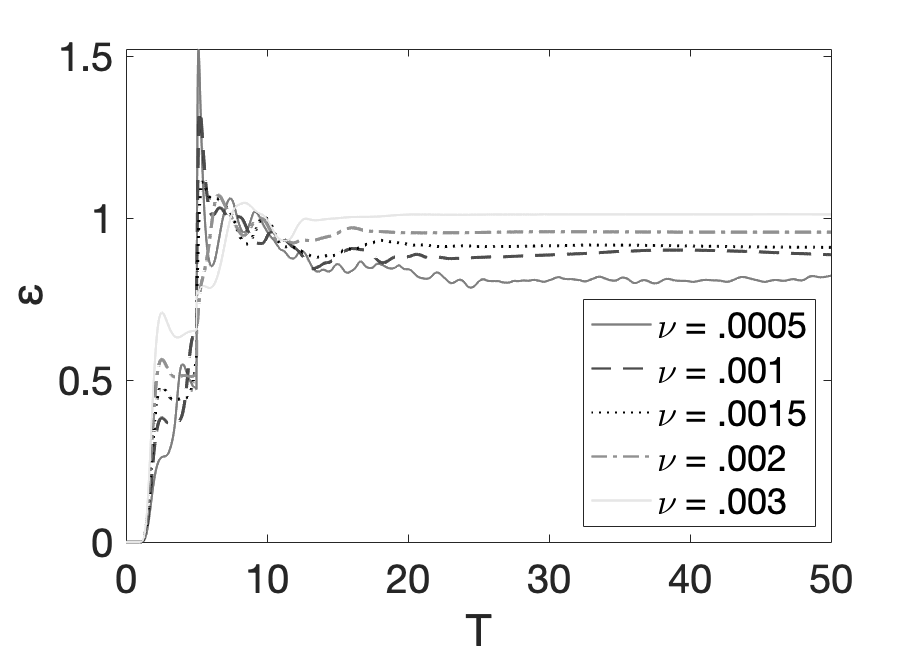}
	\caption{The energy dissipation rate over time.}
	\label{epsovertime}
\end{figure}

In Figure \ref{epsovertime} $\varepsilon (t)$ is plotted as a function of
time. The jump at $T=5$ corresponds to when the $k$ equation (and thus the
turbulent viscosity) is turned on.

To find the dependence on the Reynolds number, we plotted $\frac{%
	<\varepsilon >}{\frac{U^{3}}{L}}$ as a function of Reynolds number, and fit
to $y=a+bRe^{c}$ using Matlab's nonlinear least squares tool. The initial
guess chosen for the (iterative) solver was $y=.05+5\mathcal{R}e^{-1}.$

\begin{figure}[h]
	\centering
	\includegraphics[width=0.5\linewidth]{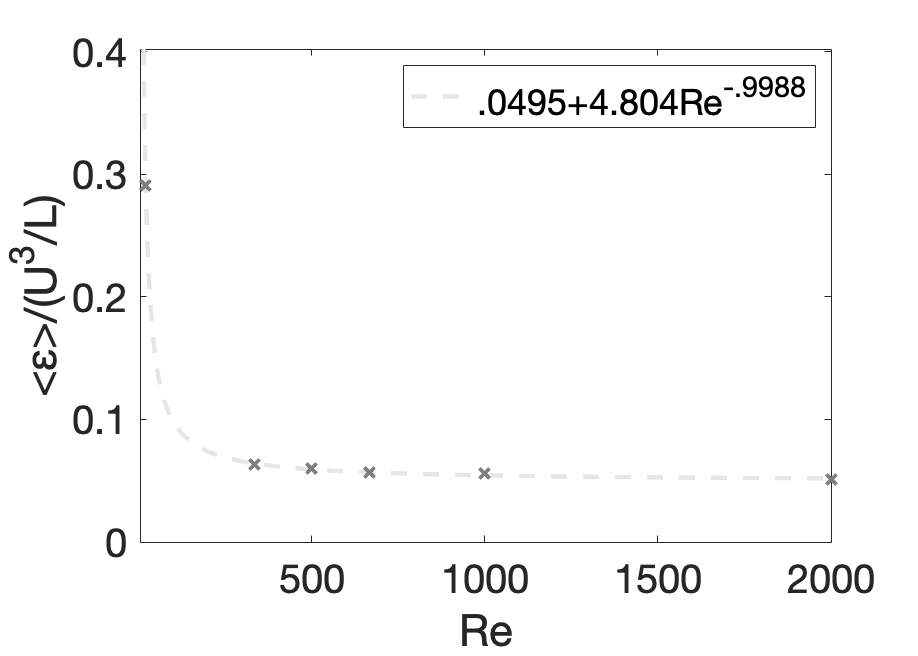}
	\caption{The energy dissipation rate over time.}
	\label{epsvsre}
\end{figure}
Figure \ref{epsvsre} shows that the long time average of the energy
dissipation rate for the model scales like a constant plus the inverse of
the Reynolds number, $\left\langle \varepsilon \right\rangle \simeq \left(
0.5+4\mathcal{R}e^{-1}\right) \frac{U^{3}}{L}$, consistent with our analysis.

\section{Conclusions and open problems}

The work herein was motivated by the idea that models more closely
reflecting the kinetic energy balance in turbulence can be simpler and
require fewer calibration parameters for accuracy. One important aspect of
kinetic energy balance is the averaged energy dissipation rate, $%
\left\langle \varepsilon \right\rangle $, in turbulence models matching
averaged energy input rates, $U^{3}/L$, as they do for the NSE. For (1.1)
this matching, related to models not over dissipating solutions, depends on
the choice of the turbulence length scale $l(\cdot )$,  the decision to include or exclude the term $-\nu
\triangle k$ in the $k-$equation and (in numerics) numerical dissipation in
the methods used. For the turbulence length scale, away from walls we used the
simple and universal kinematic specification $l=\sqrt{2}k{}^{+1/2}\tau $.
Near walls it is necessary to match the near wall behavior of $\nu
_{T}(\cdot )$ to that of the Reynolds stress $-u^{\prime }u^{\prime }$.
Including the term $-\nu \triangle k$, matching requires near wall behavior $%
l=\mathcal{O}(d^{3/2}$). With this matching, model energy dissipation rates
do match input rates, as desired for accuracy. For implementation,  $l=\min
\left\{ \sqrt{2}k{}^{+1/2}\tau ,\text{ }0.41d\sqrt{d/L}\right\} $ retains
the issue of specifying the wall distance but it does not require
pre-determining fluid sub-regions. 

The $1-$equation model studied has been used in many numerical codes, yet
open problems abound. The important analytic problems of existence and
positivity of $k$, while open for the new length scale, seem within reach
given the advances in theory presented in Chacon-Rebollo and Lewandowski \cite%
{CL}. The question of inclusion or exclusion of $-\nu \triangle k$ is little
explored. We conjecture that it is linked to the correct near wall
asymptotics of $l(\cdot )$, global dissipation rates, and possible ill-posedness of the continuum model and its numerical discretization. The model
parameters used in our tests were $\mu =0.55$ and von Karman constant $0.41$%
. These values are classical for $l=0.41d$. The numerical illustration found
that with these parameter values $\left\langle \varepsilon \right\rangle
\simeq \left( 0.5+4\mathcal{R}e^{-1}\right) \frac{U^{3}}{L}$. The $\mathcal{R%
}e\rightarrow \infty $ limiting value $0.5$ includes numerical dissipation
and grid effects. It is larger that the best estimate for the NSE of $0.088$
of Doering and Constantine \cite{DC92}. If this persists in more detailed
tests, the chosen model parameters $0.55$ and $0.41$ should be
adjusted for the new turbulence length scale herein.

When achievable, the analysis of energy dissipation rates provides a
powerful tool to investigate conditions under which turbulence models do not
severely over dissipate solutions. Naturally, the region between models
amenable to such analysis and models used in practice remains filled with
important, interesting, and challenging open problems.

\section*{Funding}
The work of the first and second author was partially supported by NSF grant DMS 1817542.

\section*{Acknowledgment}
We dedicate this paper to Charlie Doering. He was a gifted scientist and to the second author an inspiring colleague.



\section{Appendix: Existence of long time limits}

In this appendix, $C$ will denote any quantity uniformly bounded in time.
We now prove the bounds given in Section 2 on 
\begin{gather*}
	||v(T)||^{2},\int_{\Omega }k(T)dx,\int_{\Omega }\,\nu _{T}(\cdot
	,T)dx,\left\langle \frac{1}{L^{3}}\int_{\Omega }|\,\nabla ^{s}{v}%
	|^{2}dx\right\rangle _{T}, \\
	\left\langle \frac{1}{L^{3}}\int_{\Omega }\frac{1}{l}k\sqrt{k}%
	dx\right\rangle _{T}\text{ \ and \ }\left\langle \frac{1}{L^{3}}\int_{\Omega
	}[2\nu +\,\nu _{T}(\cdot )]|\,\nabla ^{s}{v}|^{2}dx\right\rangle _{T}.
\end{gather*}

\textbf{proof: \ }We begin with the energy equalities and inequalities for the two
equations:%
\begin{gather*}
	\frac{1}{2}\frac{d}{dt}||v||^{2}+\int_{\Omega }[2\nu +\,\nu _{T}(\cdot
	)]|\,\nabla ^{s}{v}|^{2}dx\leq \\
	\leq (v_{t},\phi )+\int_{\Omega }[2\nu +\,\nu _{T}(\cdot )]\nabla ^{s}{v}%
	:\nabla ^{s}\phi dx+(v\cdot \nabla v,\phi ), \\
	\text{and }\int_{\Omega }k_{t}dx+\int_{\Omega }\frac{1}{l}k\sqrt{k}%
	dx=\int_{\Omega }\nu _{T}(\cdot )|\nabla ^{s}v|^{2}dx.
\end{gather*}%
Pick $\theta ,0<\theta <1,$ and add the first equation + $\theta \times $second
equation. Using $\frac{d}{dt}||\phi ||^{2}=0$ gives%
\begin{gather}
	\frac{d}{dt}\left( \frac{1}{2}||v||^{2}-(v,\phi )+\frac{1}{2}||\phi
	||^{2}+\theta \int_{\Omega }kdx\right) +  \notag \\
	+\int_{\Omega }[2\nu +\,(1-\theta )\nu _{T}(\cdot )]|\,\nabla ^{s}{v}%
	|^{2}+\theta \frac{1}{l}k\sqrt{k}dx\leq  \label{TotalEnergy} \\
	\leq \int_{\Omega }[2\nu +\,\nu _{T}(\cdot )]\nabla ^{s}{v}:\nabla ^{s}\phi
	dx+(v\cdot \nabla v,\phi ).  \notag
\end{gather}%
We make the same choice of $\phi $ only with $\beta =\frac{1}{8}\mathcal{R}%
e^{-1}$ rather than $\beta =\frac{1}{8}\mathcal{R}e_{eff}^{-1}$. Consider
now the three terms on the RHS. For the last, nonlinear term, we have proven the
estimate%
\begin{equation*}
	\label{eqn:TotalEnergy}
	(v\cdot \nabla v,\phi )\leq C+\beta \mathcal{R}e\int_{\mathcal{S}_{\beta
	}}2\nu |\,\nabla ^{s}{v}|^{2}dx\leq C+\frac{1}{8}\int_{\Omega }2\nu
	|\,\nabla ^{s}{v}|^{2}dx,
\end{equation*}%
and the second term is subsumed in the LHS of \eqref{TotalEnergy}. The first term
on the RHS is bounded by the Cauchy-Schwarz-Young inequality in a standard
way as%
\begin{equation*}
	\int_{\Omega }2\nu \nabla ^{s}{v}:\nabla ^{s}\phi dx\leq C+\frac{1}{8}%
	\int_{\Omega }2\nu |\,\nabla ^{s}{v}|^{2}dx
\end{equation*}%
with the second term on the RHS again subsumed as above. The remaining term
on the RHS involves $\nu _{T}$. As a first step we again apply the
Cauchy-Schwarz-Young inequality in a standard way and then use the direct
calculation of $|\,\nabla ^{s}{\phi }|^{2}$ to give 
\begin{eqnarray*}
	\int_{\Omega }\,\nu _{T}(\cdot )\nabla ^{s}{v} &:&\nabla ^{s}\phi dx\leq 
	\frac{1-\theta }{2}\int_{\Omega }\nu _{T}(\cdot )|\,\nabla ^{s}{v}|^{2}dx+%
	\frac{1}{2(1-\theta )}\int_{\Omega }\nu _{T}(\cdot )|\,\nabla ^{s}{\phi }%
	|^{2}dx \\
	&\leq &\frac{1-\theta }{2}\int_{\Omega }\nu _{T}(\cdot )|\,\nabla ^{s}{v}%
	|^{2}dx+\frac{\mu }{2(1-\theta )}\left( \frac{U}{\beta L}\right) ^{2}\int_{%
		\mathcal{S}_{\beta }}l\sqrt{k}dx.
\end{eqnarray*}%
Collecting these terms in \eqref{TotalEnergy} gives%
\begin{gather}
	\frac{d}{dt}\left( \frac{1}{2}||v-\phi ||^{2}+\theta \int_{\Omega
	}kdx\right) +\int_{\Omega }\left[ \frac{3}{2}\nu +\,\frac{1-\theta }{2}\nu
	_{T}(\cdot )\right] |\,\nabla ^{s}{v}|^{2}+\theta \frac{1}{l}k\sqrt{k}dx\leq
	\notag \\
	\leq C+\frac{\mu }{2(1-\theta )}\left( \frac{U}{\beta L}\right) ^{2}\int_{%
		\mathcal{S}_{\beta }}l\sqrt{k}dx.  \notag
\end{gather}%
For the last term we apply H\"{o}lder's inequality with exponents 3 and 3/2
as follows%
\begin{gather*}
	\int_{\mathcal{S}_{\beta }}l\sqrt{k}dx=\int_{\mathcal{S}_{\beta
	}}l^{+4/3}\cdot l^{-1/3}\sqrt{k}dx\leq \left( \int_{\mathcal{S}_{\beta
	}}l^{-1}k^{+3/2}dx\right) ^{\frac{1}{3}}\left( \int_{\mathcal{S}_{\beta
	}}\left( l^{+4/3}\right) ^{3/2}dx\right) ^{\frac{2}{3}} \\
	\frac{1}{2(1-\theta )}\left( \frac{U}{\beta L}\right) ^{2}\int_{\mathcal{S}%
		_{\beta }}\mu l\sqrt{k}dx\leq \frac{1}{3}\int_{\Omega }\frac{1}{l}k\sqrt{k}%
	dx+\frac{2}{3}\left[ \frac{1}{2(1-\theta )}\left( \frac{U}{\beta L}\right)
	^{2}\right] ^{3/2}\int_{\mathcal{S}_{\beta }}l^{2}dx.
\end{gather*}%
We thus have%
\begin{gather}
	\frac{d}{dt}\left( \frac{1}{2}||v-\phi ||^{2}+\theta \int_{\Omega
	}kdx\right) +  \notag \\
	+\int_{\Omega }[\frac{3}{2}\nu +\,\frac{1-\theta }{2}\nu _{T}(\cdot
	)]|\,\nabla ^{s}{v}|^{2}+\frac{\theta }{2}\frac{1}{l}k\sqrt{k}dx\leq
	C+C^{\ast }\int_{\mathcal{S}_{\beta }}l^{2}dx,
\end{gather}%
where%
\begin{equation*}
	C^{\ast }=\frac{2}{3}\left[ \frac{1}{2(1-\theta )}\right] ^{3/2}\left( \frac{%
		U}{\beta L}\right) ^{3}.
\end{equation*}%
The result now follows by standard differential inequalities provided there
is an $\alpha >0$ with%
\begin{equation*}
	\int_{\Omega }\frac{1}{l}k\sqrt{k}dx\geq \alpha \int_{\Omega }kdx\text{ \
		and \ }\int_{\mathcal{S}_{\beta }}l^{2}dx\leq C<\infty .
\end{equation*}%
These two depend on the choice of $l=\min \left\{ \sqrt{2}k{}^{+1/2}\tau ,%
\text{ }0.41d\sqrt{\frac{d}{L}}\right\} $. By selecting the last argument in
the minimum, the condition $\int l^{2}dx\leq C<\infty $ holds. By selecting
the first term in the minimum (and noting that then $\frac{1}{l}k\sqrt{k}=%
\frac{1}{\sqrt{2}\tau }k$) the condition $\int \frac{1}{l}k\sqrt{k}dx\geq
\alpha \int kdx$\ holds. Thus the uniform bounds follows.

\end{document}